\documentclass{article}
\usepackage{spconf,amsmath,graphicx}
\usepackage{multirow}
\usepackage{bm}
\newcommand{\etal}{\textit{et~al.} }
\usepackage[caption=false, font=footnotesize]{subfig}

\usepackage[firstpage=true]{background}
\SetBgContents{%
	\fbox{%
		\parbox{\textwidth}{%
			\footnotesize \textcopyright 2024 IEEE. Personal use of this material is permitted. Permission from IEEE must be obtained for all other uses, in any current or future media, including reprinting/republishing this material for advertising or promotional purposes, creating new collective works, for resale or redistribution to servers or lists, or reuse of any copyrighted component of this work in other works. DOI: https://doi.org/10.1109/ICIP51287.2024.10647966. }}%
}	
\SetBgScale{1}
\SetBgAngle{0}
\SetBgPosition{current page.north}
\SetBgVshift{-1.6cm}
\SetBgColor{black}
\SetBgOpacity{1}

\def\ninefivept{\def\baselinestretch{1.11}\let\normalsize\small\normalsize}

\title{Efficient Learned Wavelet Image and Video Coding}

\name{Anna Meyer$^{\star}$ \qquad Srivatsa Prativadibhayankaram$^{\star \dagger}$ \qquad Andr\'e Kaup$^{\star}$\thanks{The authors gratefully acknowledge that this work has been funded by the Deutsche Forschungsgemeinschaft (DFG, German Research Foundation) under project number 461649014.}}
\address{$^{\star}$ \textit{Multimedia Communications and Signal Processing},\\ Friedrich-Alexander-University Erlangen-Nürnberg,
	Erlangen, Germany\\
	$^{\dagger}$ \textit{Moving Picture Technologies},\\
	Fraunhofer Institute for Integrated Circuits, Erlangen, Germany }

\begin{document}
%
\maketitle
\begin{abstract}
Learned wavelet image and video coding approaches provide an explainable framework with a latent space corresponding to a wavelet decomposition. The wavelet image coder iWave++ achieves state-of-the-art performance and has been employed for various compression tasks, including lossy as well as lossless image, video, and medical data compression. However, the approaches suffer from slow decoding speed due to the autoregressive context model used in iWave++. In this paper, we show how a parallelized context model can be integrated into the iWave++ framework. Our experimental results demonstrate a speedup factor of over 350 and 240 for image and video compression, respectively. At the same time, the rate-distortion performance in terms of Bj{\o}ntegaard delta bitrate is slightly worse by 1.5\% for image coding and 1\% for video coding. In addition, we analyze the learned wavelet decomposition by visualizing its subband impulse responses.
\end{abstract}

\begin{keywords}
	Learned image compression, learned video compression, discrete wavelet transform, interpretability 
\end{keywords}
	\vspace{-0.2cm}
\section{Introduction}
\label{sec:intro}
Until the early 2000s, image and video compression based on discrete wavelet transforms has been a highly active field of research. The wavelet transform has desirable properties for compression because its compromise between spatial and frequency resolution fits the correlation structure of images. With the emergence of predictive transform coding as dominant principle in video coding, wavelet-based methods have received less attention. Similarly, the main paradigm in learned image and video compression is nonlinear transform coding \cite{Balle2021}. 

However, learned wavelet transforms have recently been employed for image \cite{maiwave++, Xue2023a} and video \cite{Meyer2023} compression. The approaches achieve state-of-the-art performance, which demonstrates the potential of learned wavelet transforms. Wavelet-based frameworks provide an explainable coding scheme, as their latent space has a defined structure corresponding to a wavelet decomposition. In addition, learned wavelet coding approaches support lossless compression, which makes the wavelet coding scheme well suited for specialized tasks such as medical image and video compression \cite{Xue2021, Xue2023, Xue2023b}.

In this paper, we address a limitation of the current learning-based wavelet coding frameworks in the literature: Their backbone is the image coder iWave++ \cite{maiwave++}, which relies on an autoregressive context model. Because of slow sequential decoding, a more efficient learned wavelet coding approach is required. We propose a parallel context model suited for the subband latent space and show how it can be integrated into the iWave++ framework. Our experiments demonstrate that our approach can be used for image as well as video coding models, achieving a significant decoding speedup at a small loss in terms of rate-distortion performance. 
	\vspace{-0.4cm}
\section{Related Work}
\label{sec:related}
\subsection{Learned Wavelet Image and Video Coding}
	\vspace{-0.1cm}
Ma \etal \cite{maiwave++} introduced an end-to-end optimizable wavelet image coding framework called iWave++. The analysis and synthesis transforms correspond to a trainable wavelet transform \cite{Ma2020} implemented using the lifting scheme. With the wavelet transform based on Convolutional Neural Networks (CNNs), iWave++ is competitive to state-of-the-art image coders including VVC. Xue \etal \cite{Xue2023a} improved the performance of iWave++ with a framework called iWavePro. They extended the context model to capture dependencies between color components, similar as in \cite{icassp2023}. Furthermore, Xue \etal incorporated an affine wavelet transform module and an online training strategy that adapts to the image to be coded during inference. 

The iWave++ framework has also been employed for video coding: Dong \etal \cite{Dong2022} introduced a partly trainable wavelet video coder. The coder first performs a traditional non-trainable temporal wavelet transform. Next, the temporal subbands are coded using an entropy parameter estimation module that follows the structure of iWave++. In \cite{Meyer2023}, an end-to-end trainable wavelet video coder was proposed. Here, the input video sequence is processed by a temporal wavelet transform implemented using a trainable motion compensated temporal filtering (MCTF) module. The obtained temporal subbands are subsequently compressed by iWave++.

The approaches described above have one limitation in common: To achieve state-of-the-art performance, they rely on the iWave++ framework including an autoregressive context model. As the learned wavelet transform does not reduce the number of symbols in the transform space, this leads to prohibitively long decoding times. 
	\vspace{-0.3cm}
\subsection{Efficient Context Models}
	\vspace{-0.1cm}
An autoregressive context model implemented using 2D masked convolutions significantly improved the rate-distortion performance of learned image coders \cite{minnen2018joint}, but requires sequential decoding. Context models that can be parallelized are hence required to accelerate decoding speed and different parallelization techniques have been investigated in the literature. Minnen \etal \cite{Minnen2020} group the channels of the latent space into a fixed number of chunks. They code these chunks sequentially enable conditioning on previously coded channels. A checkerboard context model \cite{He2021} allows conditioning on spatial context: The black positions of a checkerboard are coded in a first coding step and provide context for the white positions coded in a second coding step. He \etal \cite{He2022} further developed this approach by combining the checkerboard model with channel-wise conditioning. They divide the latent space channels into uneven groups, motivated by the varying information content of each channel. In \cite{Koyuncu2023}, Koyuncu \etal also utilize a combination of spatial and channel-wise grouping mechanisms to parallelize a transformer-based context model. Li \etal \cite{Li2023} use a four-step spatial context model combined with channel conditioning for their video compression model. 

Because the latent space of the wavelet image coding framework iWave++ corresponds to a wavelet decomposition, channel-wise conditioning cannot be exploited for the single-channel subbands. Nevertheless, our proposed four-step context model achieves similar rate-distortion performance compared to the autoregressive wavelet coder with spatial conditioning alone. 
\section{Proposed Method}
	\vspace{-0.3cm}
We first provide an overview of our proposed parallelized wavelet image coder, which we refer to as pWave++. Then, we introduce a corresponding wavelet video coder. Also, we discuss a visualization technique that allows comparing the wavelet transform to other transforms in learned compression.
\subsection{Image Compression Model: pWave++}
\begin{figure}[tb]
	\centering	
	\includegraphics[width=0.49\textwidth]{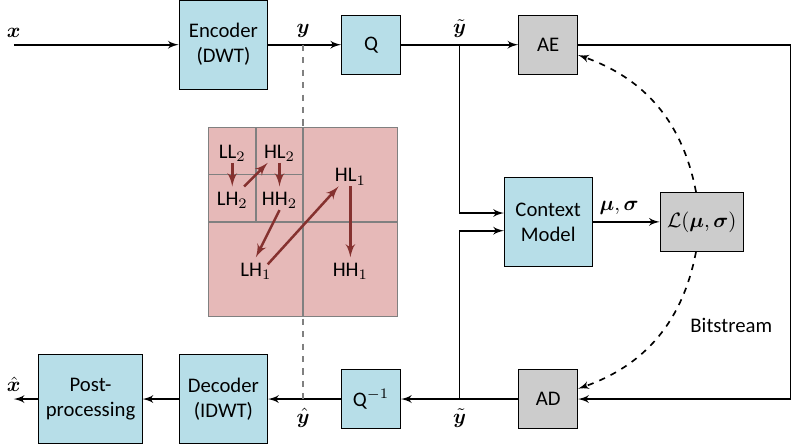}
	\caption{Overview of the end-to-end wavelet image coding framework iWave++ \cite{maiwave++}. All blue modules are trainable. The input $\bm{x}$ is a color component of an input image in the YCbCr color space. The subbands $\bm{y}$ obtained from the discrete wavelet transform (DWT) are processed in the coding order visualized by the red arrows. }.
	\label{fig:iwave}
	\vspace{-0.7cm}
\end{figure}
\begin{figure}[tb] \hspace{-0.85cm}
	\includegraphics[width=0.6\textwidth]{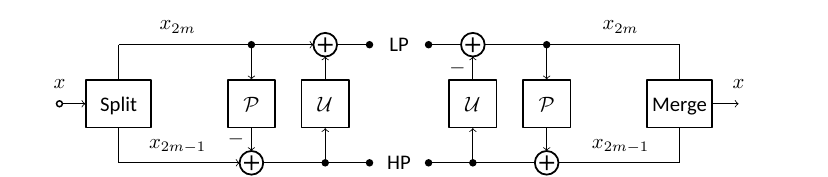} 	
	\caption{The one dimensional forward and inverse lifting scheme. Perfect reconstruction is guaranteed if the output of the predict $\mathcal{P}(\cdot)$ and update $\mathcal{U}(\cdot)$ filters is integer-valued.}
	\label{fig:lifting}
		\vspace{-0.5cm}
\end{figure}
\begin{figure*}[tb]
	\centering	
	\includegraphics[width=0.8\textwidth]{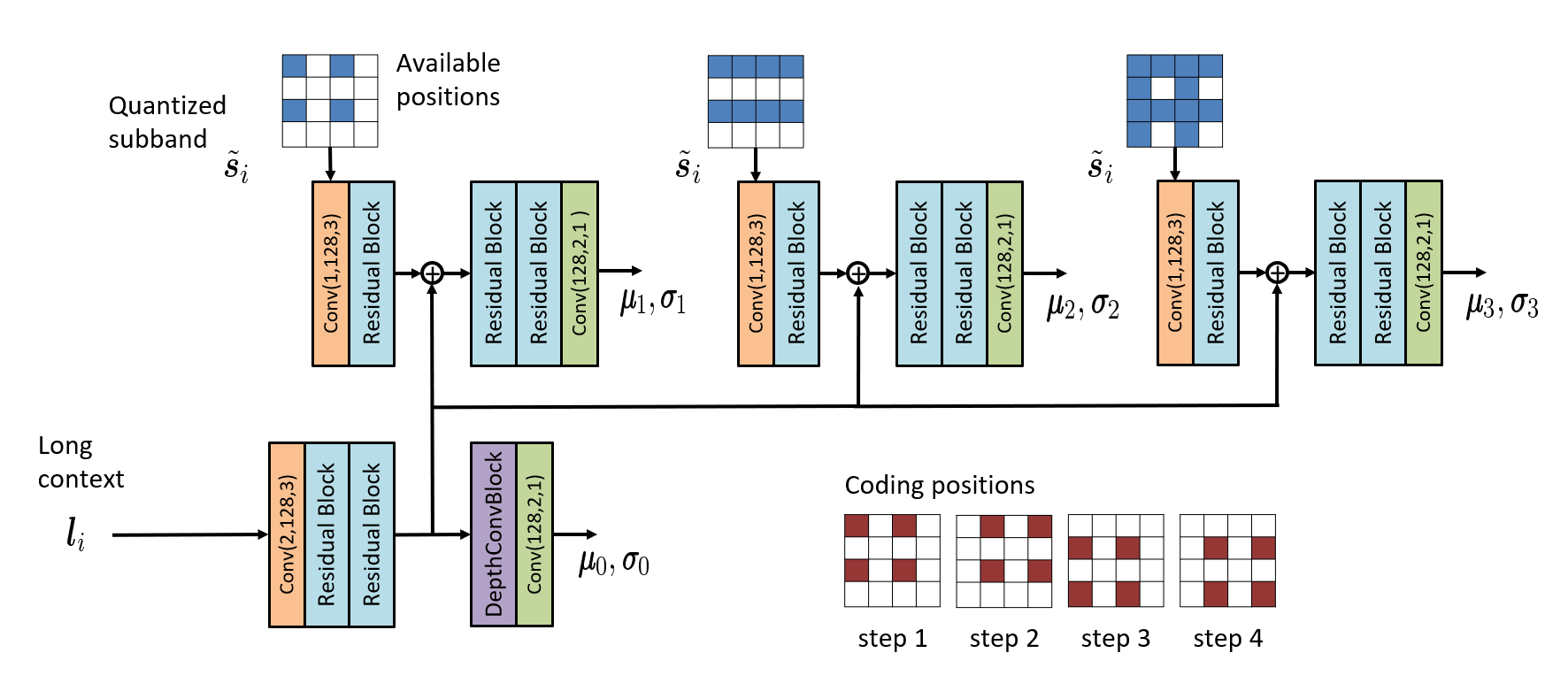}
	\caption{Proposed four-step context fusion model. Convolutional layers are specified as \textit{Conv}(input channels, output channels, kernel size). The positions to be coded and the available previously coded symbol positions are visualized for each of the four coding steps. The detailed architecture of the \textit{DepthConvBlock}  and \textit{Residual Block} modules  are provided in Fig.~\ref{fig:contextfusionb}. }
	\label{fig:contextfusion}
	\vspace{-0.4cm}
\end{figure*}
\begin{figure}[tb]
	\centering	
		\vspace{-0.2cm}
	\subfloat[\footnotesize LL subband $\tilde{\bm{s}}_1$]{%
		\hspace{1.8cm}	\includegraphics[width=0.3\textwidth]{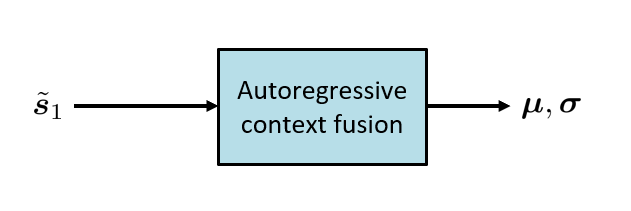}
	}\\ 
	\vspace{-0.4cm}	
	\subfloat[\footnotesize Remaining subband types $\tilde{\bm{s}}_i, i=2, \hdots, 13$ ]{%
		\includegraphics[width=0.38\textwidth]{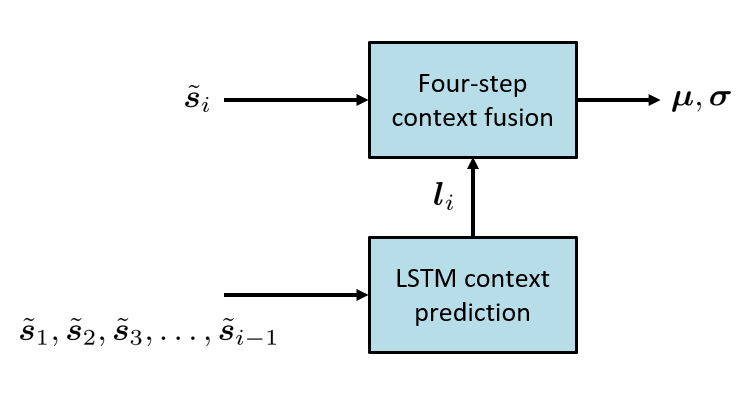}
	}
	\caption{The context model of pWave++ for entropy parameter estimation depending on the subband type. }
	\label{fig:contextmodel}
	\vspace{-0.5cm}
\end{figure}

Our pWave++ model follows the structure of the end-to-end wavelet image coding framework iWave++. Fig.~\ref{fig:iwave} provides an overview of the approach. The input $\bm{x}$ is one color component of an image in the YCbCr color space. The analysis transform corresponds to a learned discrete wavelet transform (DWT) and the synthesis transform to the inverse discrete wavelet transform (IDWT). The latent space $\bm{y}$ hence has a known structure and represents a wavelet decomposition. The subband layout is shown in the middle of Fig.~\ref{fig:iwave} with two decomposition levels for visualization. Our model has four decomposition levels in total. The subbands $\bm{y}$ are quantized by scalar quantization according to $\tilde{\bm{y}} =\lfloor \bm{y} \cdot \Delta \rceil $, where straight-through gradients are used for the rounding operation during training. At the decoder side, the reconstructed subbands are obtained as $\hat{\bm{y}}  = \tilde{\bm{y}} / \Delta $. We use a dedicated trainable parameter $\Delta_{\mathrm{L}}$ for the lowpass subband and another parameter $\Delta_{\mathrm{H}}$ for the remaining subband types, in contrast to the baseline model with a single parameter for all subbands. As indicated by the red arrows in Fig.~\ref{fig:iwave}, the quantized subbands $\tilde{\bm{y}}$ are coded sequentially. The model uses adaptive arithmetic coding to compress the quantized wavelet representation. The context model for entropy parameter estimation will be explained in the following Section~\ref{sec:contextmodel}. We employ a Laplacian distribution as probability model following \cite{Li2023}. After the inverse wavelet transform, the final reconstruction $\hat{\bm{x}}$ is obtained from a post-processing module. Because the encoding and decoding transform share their parameters, this module is required to compensate quantization artifacts.

The wavelet transform is implemented using the lifting scheme, which allows constructing second generation wavelets \cite{sweldens1995NewPhil}. The 1D lifting scheme shown in Fig.~\ref{fig:lifting} consists of the three steps split, predict, and update. The signal $x$ is split into even- and odd-indexed samples. A highpass (HP) representation is obtained from the predict operation as $\mathrm{HP} = x_{2m-1} - \mathcal{P}(x_{2m})$, predicting odd from even samples. With the update operation, global properties of the signal are maintained and a lowpass (LP) representation is obtained as $\mathrm{LP} = x_{2m} + \mathcal{U}(\mathrm{HP})$. For a 2D wavelet decomposition, the 1D lifting scheme is first applied row-wise and then column-wise. The trainable transform is realized by inserting residual CNN modules as prediction and update filters. Details on the filter architecture can be found in \cite{maiwave++, icassp2023}.
	\vspace{-0.3cm}
\subsection{Parallel Context Model}
\label{sec:contextmodel}
The context model of pWave++ is purely backward adaptive, exploiting spatial context from the current subband to be coded as well as hierarchical context from previously coded subbands. As the LL subband is the first subband to be coded and one of the subbands with the smallest resolution ($H/16\times W/16\times 1$), we adopt the autoregressive context fusion model from iWave++ (architecture details are provided in \cite{maiwave++, icassp2023}).  As can be seen in Fig.~\ref{fig:contextmodel}(a), the means and scales for the Laplace distribution are estimated based on spatial context from the LL subband $\tilde{\bm{s}}_1$ only. The remaining subband types are coded utilizing context from previously coded subbands combined with spatial context, as shown in Fig.~\ref{fig:contextmodel}(b). A context fusion module combines causal spatial context from $\tilde{\bm{s}}_i$ with the long term context $\bm{l}_i$. The long term context is a hidden state of a convolutional Long-Short Term Memory (LSTM) unit from the context prediction module. The LSTM-based module facilitates exploiting dependencies from previously coded subbands. Contrary to the findings in \cite{Xue2023a}, we found that the context prediction module stabilizes training in our experiments. 

The architecture of the four-step context fusion model inspired by \cite{Li2023} is depicted in Fig.~\ref{fig:contextfusion}. Instead of sequential decoding with the autoregressive model, the proposed context fusion model only requires four coding steps independent of the subband size. In the first coding step, the entropy parameters $\bm{\mu}_0, \bm{\sigma}_0$ are estimated based on the long context $\bm{l}_i$. Here, only one fourth of the pixel positions are coded, defined by a wide checkerboard pattern shown at the bottom right of Fig.~\ref{fig:contextfusion}. Those symbol positions are subsequently available for the second coding step (cf. top left of Fig.~\ref{fig:contextfusion}). The previously coded coefficients from the current subband $\tilde{\bm{s}}_i$ are combined with the long context to estimate $\bm{\mu}_1, \bm{\sigma}_1$ for the second coding step. In the third coding step, half of the coefficients from the current subband are available to estimate $\bm{\mu}_2, \bm{\sigma}_2$ in the same manner as in the previous step. The fourth coding step codes the remaining positions by again combining the causal context from $\tilde{\bm{s}}_i$ with the long context to estimate $\bm{\mu}_3, \bm{\sigma}_3$. Note that the described four-step approach on average provides four neighbors for a $3\times3$ kernel (0, 2, 6, and 8 neighbors in the respective coding steps). The full autoregressive model on average has access to four neighbors, too.

\begin{figure}[tb]	
	\centering	
	\includegraphics[width=0.365\textwidth]{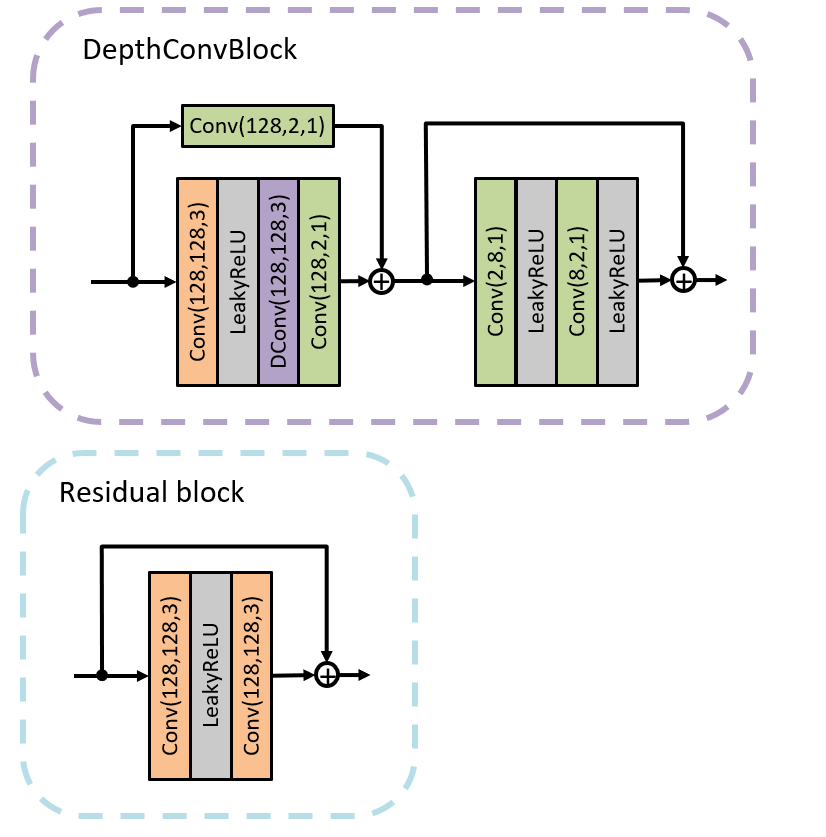}
	
	\caption{Architecture details of the building blocks used in the four-step context fusion model. Convolutional layers are specified as \textit{Conv}(input channels, output channels, kernel size).  }
	\label{fig:contextfusionb}
	\vspace{-0.3cm}
\end{figure}

The two main building blocks of the proposed context fusion model are illustrated in Fig.~\ref{fig:contextfusionb}. The \textit{DepthConvBlock} at the top is a two-part residual block and is exclusively used as building block by \cite{Li2023}. The first part of the residual structure contains a depth-wise convolution, i.e., a regular convolution with 128 groups that convolves each input channel with a separate filter kernel. The second main building block of our model (bottom of Fig.~\ref{fig:contextfusionb}) is a two-layer \textit{Residual Block}. In our experiments, we found that the combination of using a \textit{DepthConvBlock} in the first coding step and \textit{Residual Blocks} in the remaining steps performed best. In particular, the observed rate-distortion performance was significantly better compared to using either only \textit{DepthConvBlocks} or \textit{Residual Blocks} for all four coding steps.
 		\vspace{-0.3cm}
\subsection{Parallel Wavelet Video Coder: pMCTF}
	\vspace{-0.1cm}
\begin{figure}[tb]
	\centering	
	\includegraphics[width=0.49\textwidth]{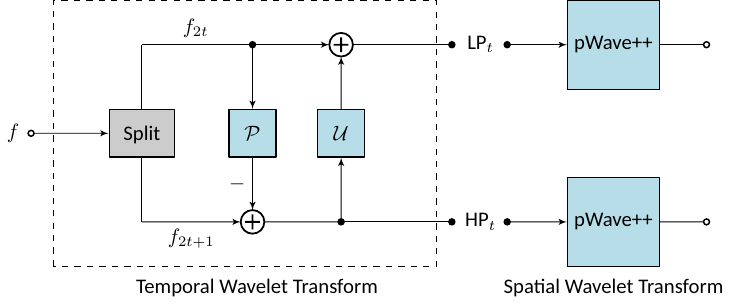}
	\caption{Overview of our video coding framework pMCTF. \quad $f$ denotes the input video sequence.}
	\label{fig:mctf}
			\vspace{-0.3cm}
\end{figure}
We use our proposed image coding model to build a faster wavelet video coder based on MCTF \cite{Meyer2023}. Fig.~\ref{fig:mctf} provides an overview of the proposed pMCTF framework.
The temporal wavelet transform is realized using MCTF. It first computes a temporal low- and highpass subband using trainable predict and update steps. The predict step performs a motion-compensated prediction, while the update step re-uses the computed motion vectors to perform inverse motion compensation on the highpass subband. The update step thereby performs lowpass filtering along the motion trajectory and is essential for the coding efficiency of learned MCTF. After the temporal wavelet transform, the low- and highpass subbands are coded independently by pWave++ models. 
For details about the MCTF framework, the reader is referred to \cite{Meyer2023}.
		\vspace{-0.3cm}
\subsection{Subband Impulse Response}
\label{sec:sir}
The channel impulse response visualization can be used as a tool for a better understanding of the latent space features of compressive autoencoder architectures \cite{Duan2022, PV2023a}. We adapt the approach to obtain subband impulse responses for the learned nonlinear wavelet transform, allowing a comparison to other transforms in learned compression.

First, we take one color component of an image in the YCbCr color space and perform the learned wavelet transform. Then, we obtain the subband impulse response for each of the 13 subbands as follows: For every quantized subband $\tilde{\bm{s}}_i$ with index $i=1, \hdots, 13$ we select the maximum absolute value, as the wavelet coefficients can take positive or negative values. We initialize all-zero subbands, where the subbands in the fourth decomposition level have a resolution of $1\times1\times1$ (height$\times$width$\times$channels). Thus, we have an overall resolution of $16\times16\times1$. We place the maximum absolute value in the $i$-th all-zero subband. If necessary, we obtain its location in the respective subband using max pooling. Finally, we apply the inverse learned wavelet transform and obtain a subband impulse response of size $16\times16\times1$. 
		\vspace{-0.3cm}
\section{Experiments and Results} 
\label{sec:experiments}
		\vspace{-0.1cm}
\subsection{Experimental Setup and Training}
We follow the training conditions from \cite{icassp2023} and \cite{Meyer2023} for the image model pWave++ and video model pMCTF, respectively. That way, we ensure a fair comparison to the baseline models. As training data for both the image and video models, we use the Vimeo90K \cite{xue2019video} data set. We use AdamW as optimizer, train on the luma channel only, and use the same five rate-distortion tradeoff parameters $\lambda =  \left\{ 0.007, 0.01, 0.03, 0.05, 0.08 \right\}$ for both model types. For image coding, we use randomly cropped $256\times256$ patches and for video coding, the patch size is $128$. We train the image model pWave++ from scratch for 27 epochs with the standard rate-distortion loss $\mathcal{L} =	R + \lambda \cdot  D_{\mathrm{MSE}}$. The multi-stage training strategy for pMCTF is described in \cite{Meyer2023}. For both the image and video models, we train one model with $\lambda=0.08$ and use the checkpoint to finetune for the remaining $\lambda$ values. For the image models, we allow finetuning for up to 10 epochs, while 2 epochs are sufficient for pMCTF.
		\vspace{-0.2cm}
\subsection{Image Coding: pWave++}
		\vspace{-0.1cm}
\begin{figure}[tb]
	\centering	
	\includegraphics[width=0.35\textwidth]{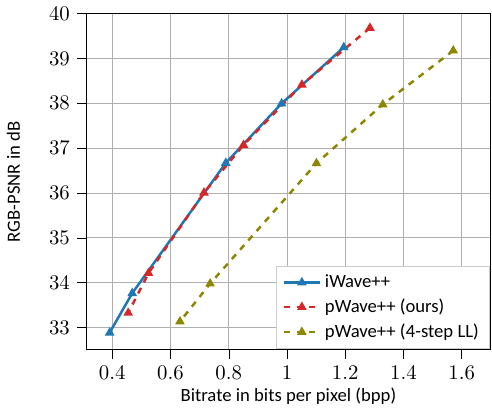}	
	\vspace{-0.3cm}  
	\caption{Rate-distortion performance on the Kodak data set.}
	\label{fig:kodak}
	\vspace{-0.3cm}
\end{figure}

Our pWave++ model achieves similar rate-distortion performance compared to the baseline iWave++ model, as shown in Fig.~\ref{fig:kodak}.
We measure a decreased coding efficiency with an increased Bj{\o}ntegaard delta (BD) bitrate of +1.5~\% in comparison to the baseline model. Note that this deterioration is caused by the lowest rate point of pWave++ alone, as its rate-distortion curve otherwise matches the curve of iWave++ (cf. Fig.~\ref{fig:kodak}). Table~\ref{tab:time} provides runtime measurements comparing the baseline model iWave++ to our proposed pWave++. The runtimes are averaged over all images from the Kodak data set and performed on a NVIDIA RTX A6000 GPU. Both the encoding and decoding time significantly decrease with our pWave++ model. We measure around 6 seconds decoding time instead of 1995 seconds, which corresponds to a speedup factor of 354.

\begin{table}[tb]
	\centering
	\caption{ Runtime measurements performed on a NVIDIA RTX A6000 GPU. We compare the baseline models to our pWave++ and pMCTF. The measurements are based on the luma component and specified in seconds. For the image models, we provide average runtimes for the Kodak data set (resolution $512 \times 768$). For the video models, we center crop patches of size $512 \times 768$ from the UVG sequences.}
	\vspace{0.5mm}
	\resizebox{0.5\textwidth}{!}{
			\begin{tabular}{l|c|c|c}
				\multirow{2}{*}{Model} & Encoding  & Decoding  & Decoding \\
				& time & time  & speedup \\
				 \hline
				iWave++ &  3.17~s & 1994.51~s & - \\
				pWave++ (ours) & 0.35~s & 5.64~s& 354 \\
				pWave++ (4-step LL)& 0.36~s & 0.35~s & 5699\\
				\hline \hline
				MCTF &  1.78~s & 1865.61~s & - \\
				pMCTF (ours) & 0.56~s & 7.656~s & 244\\
				\end{tabular}
	}
		\vspace{-0.5cm}
	\label{tab:time}
\end{table}
Using the autoregressive context fusion model for the first subband to be coded, i.e., the subband LL$_4$, is important for the rate-distortion performance of pWave++. We demonstrate this by training a pWave++ model with a four-step context fusion model for the LL$_4$ subband. As can be seen in Table~\ref{tab:time}, the full four-step model achieves an even faster decoding time. Omitting the autoregressive model leads to faster decoding by a factor of 16. However, the rate-distortion performance of the full four-step model is significantly worse, as can be seen from the rate-distortion curves in Fig.~\ref{fig:kodak}. This behavior is caused by a different subband layout of the full four-step pWave++ model: Coding the LL$_4$ subband without autoregressive spatial context lets the model store a lowpass representation in the subbands LH$_1$, HL$_1$, and HH$_1$, and creates a highpass representation towards the lower decomposition levels, with the sparsest subbands being LL$_4$, LH$_4$, HL$_4$, and HH$_4$. This finding highlights that the subband layout known from the traditional wavelet transform - with a lowpass representation in the  LL$_4$ subband and with sparse directional details in the remaining subbands - is crucial for the performance of iWave++ as well as pWave++.

\begin{figure}[tb]
	\centering	
	\includegraphics[width=0.36\textwidth]{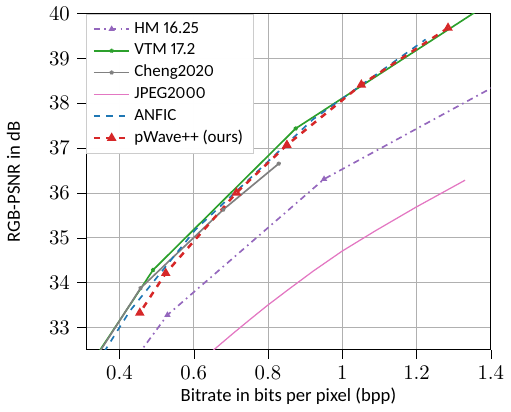}	 
	\vspace{-0.26cm}
	\caption{Rate-distortion performance on the Kodak data set. }
	\label{fig:kodak2}
	\vspace{-0.3cm}
\end{figure}
In Fig.~\ref{fig:kodak2}, we provide a comparison of our method to other state-of-the-art image coders. The rate-distortion curves of the learning-based coders Cheng2020 \cite{Cheng2020} and ANFIC \cite{Ho2021b} show that they are competitive to VTM 17.2 (green curve). Our model (red curve) shows similar rate-distortion performance compared to VTM as well. Notably, the learned wavelet image coder performs much better than the traditional wavelet-based JPEG2000.
		\vspace{-0.3cm}
\subsection{Video Coding: pMCTF}
\begin{figure}[tb]
	\centering	
	\includegraphics[width=0.35\textwidth]{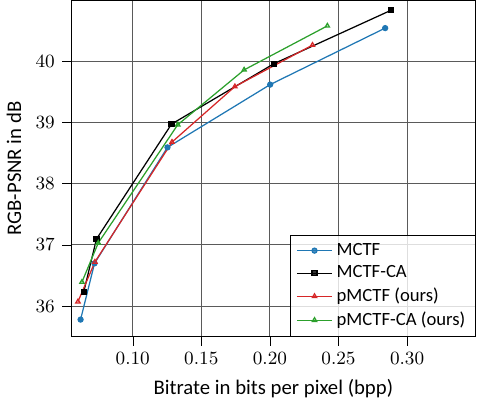}	  
		\vspace{-0.4cm}
	\caption{Rate-distortion performance on the UVG data set.   }
	\label{fig:uvg}
		\vspace{-0.1cm}
\end{figure}

		\vspace{-0.1cm}
We follow the test conditions from \cite{Meyer2023} and compare different video coders on the first 96 frames of each of the seven sequences from the UVG \cite{Mercat2020} data set. We compare our proposed pMCTF model to the baseline autoregressive MCTF model from \cite{Meyer2023} with a GOP size of 8. The corresponding rate-distortion curves can be seen in Fig.~\ref{fig:uvg}. Our pMCTF (red curve) achieves BD rate savings of -3.04\% compared to the baseline (blue curve) due to improved performance at higher rates. pMCTF decreases the average decoding time from 1866 seconds to around 8 seconds, achieving a speedup by a factor of 243 (cf. Table~\ref{tab:time}). Because the MCTF coding scheme with multiple temporal decomposition levels is sensitive to the GOP size choice, we also evaluate the content adaptive option MCTF-CA \cite{Meyer2023}. The black and green rate-distortion curves in Fig.~\ref{fig:uvg} show the baseline MCTF-CA model and our faster pMCTF-CA model, respectively. The BD rate of pMCTF-CA is slightly worse by 0.97\%. However, the parallel model still performs better at higher rates.

\begin{figure}[tb]
	\centering	
	\includegraphics[width=0.35\textwidth]{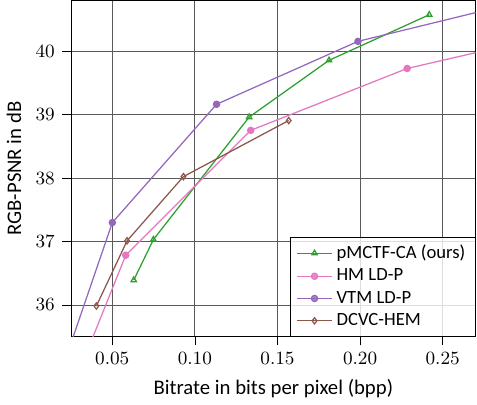}	  
	\vspace{-0.4cm}
	\caption{Rate-distortion performance on the UVG data set.  }
	\label{fig:uvg2}
		\vspace{-0.4cm}
\end{figure}

Fig.~\ref{fig:uvg2} compares our approach to state-of-the-art video coders using the same test conditions as above. We use HM 16.25 and VTM 17.2 in Lowdelay P (LD-P) configuration. The learning-based approach DCVC-HEM \cite{Li2022} shows better rate-distortion performance at lower rates and outperforms HM for a GOP size of 8. Our pMCTF-CA model performs particularly well at higher rates due to its capability of lossless compression. Here, our model surpasses VTM for bitrates larger than $\approx$ 0.22 bpp, but VTM outperforms all other approaches in a wide bitrate range.
		\vspace{-0.3cm}
\subsection{Subband Impulse Response}
		\vspace{-0.1cm}
\begin{figure}[tb]
		\vspace{-0.2cm}
	\centering	
	\subfloat{%
		\hspace{0.48cm}	\includegraphics[width=0.18\textwidth]{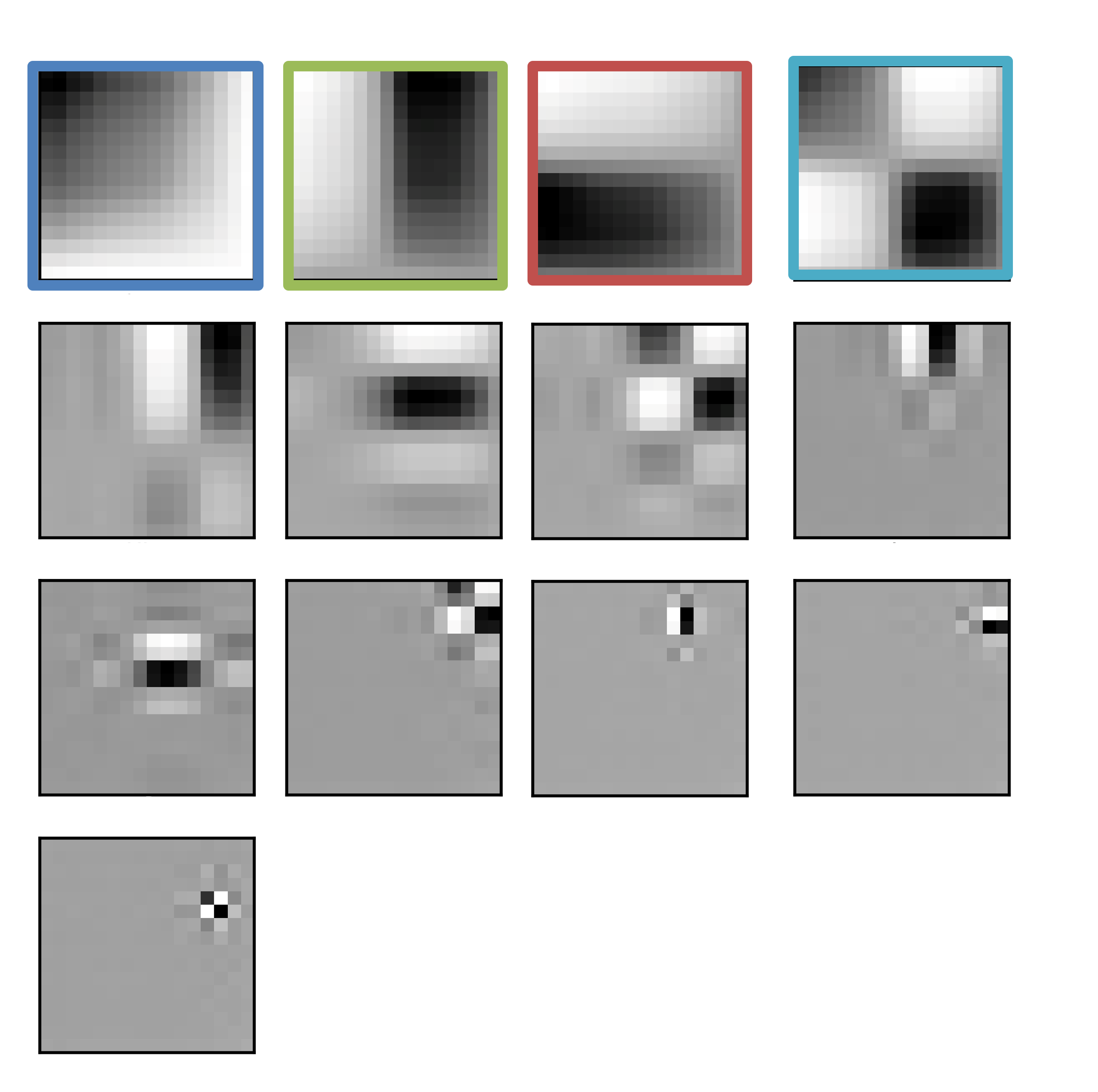}
	}\\ 
	\vspace{-0.3cm}	
	\subfloat{%
		\includegraphics[width=0.41\textwidth]{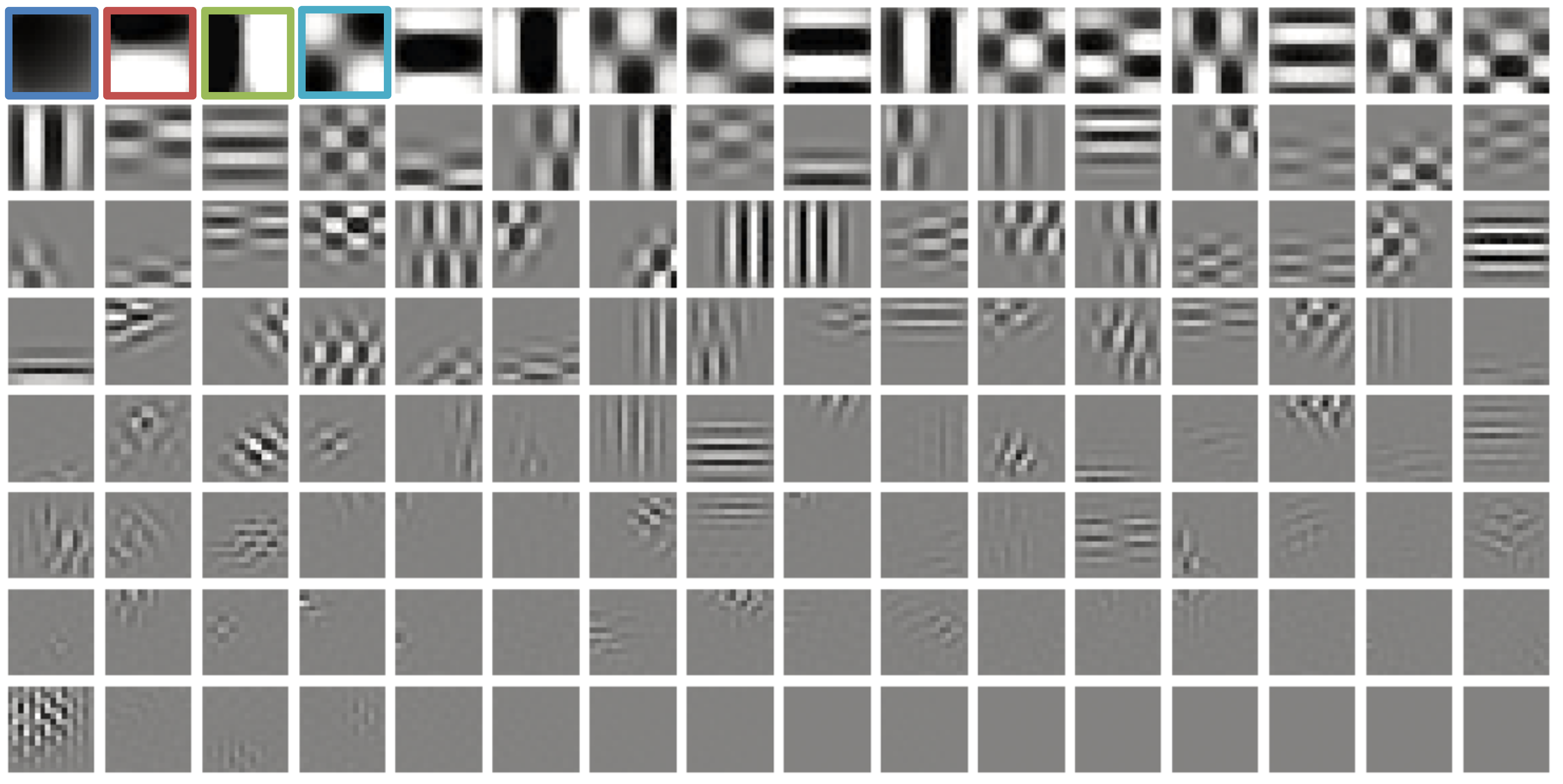}
	} 	\vspace{-0.2cm}
	\caption{Luminance impulse responses for the ''bike'' image from the JPEGXL data set. Top: Our pWave++ model. Subband impule responses ordered according to subband order $LL_4, HL_4, \hdots, HH_1$. Bottom: Color separation model \cite{PV2023a}. Channel impulse responses arranged in decreasing order of channel bitrate contribution. }
	\label{fig:impulse}
		\vspace{-0.5cm}
\end{figure}
As described in Section~\ref{sec:sir}, we compute the luminance subband impulse responses of our pWave++ model and compare the results to the channel impulse responses of the color separation model \cite{PV2023a}. To this end, we use the ''bike''\footnote{https://github.com/libjxl/conformance/blob/master/testcases/bike/ref.png} image from the JPEGXL test set. We can see that the subband impulse responses at the top of Fig.~\ref{fig:impulse} closely resemble the two-dimensional basis functions of a Haar wavelet transform. The learned wavelet transform of our model therefore captures similar structures as a traditional wavelet transform, but allows for optimization in a rate-distortion sense. Joint training with a learned entropy model enables learning-based wavelet coders to adapt to image statistics and thus outperform traditional wavelet coders like JPEG2000. When comparing the subband impulse responses to the channel impulse responses at the bottom of Fig.~\ref{fig:impulse}, it can be seen that some latent space channels respond to the same structures as specific subbands. In particular, exact matches can be found for the lowest decomposition level, which is highlighted by colored boxes. Many of the remaining channel impulse responses show localized frequency responses, which are similar to the subband impulse responses. 
\vspace{-0.4cm}
\section{Conclusion}
\label{sec:conclusion} 		\vspace{-0.25cm}
In this paper, we introduced a novel parallelized context model for the image coder iWave++ that achieves a decoding speedup by a factor of over 350 for image coding and of over 240 for video coding. Our experimental results demonstrated that the efficient model can be employed for wavelet image and video coding with small BD rate losses of up to 1.5\%. We showed that the interpretable subband layout known from traditional wavelet transforms performs best and provided a visual analysis in comparison to other transforms in learned compression. In future work, we will focus on improving the prediction capabilities of our pMCTF model by investigating advanced methods for motion estimation and compensation or integrating multiple reference frames in MCTF.  
		\vspace{-0.3cm}

\ninefivept
\bibliographystyle{IEEEbib}
\bibliography{wavelets}

\end{document}